\documentclass{IEEEtran}
\usepackage{graphicx}
\usepackage{amsmath,amssymb,amsthm}
\usepackage{mathtools}
\usepackage{float}
\usepackage{xcolor,colortbl}
\usepackage{color}
\usepackage{fp}
\usepackage{tikz}
\usetikzlibrary{shapes.misc,shadows}
\usepackage{multirow}
\usepackage{xcolor,colortbl}
\usepackage{caption}
\usepackage{graphicx}
\usepackage[linesnumbered,ruled]{algorithm2e}
\usepackage{epstopdf}
\usepackage{pgfplots}
\pgfplotsset{compat=1.7}
\usepackage{caption}
\usepackage{subcaption}
\newtheorem{lemma}{Lemma}

\usepackage{url}
\usepackage[noadjust]{cite}
\usepackage{setspace}
\graphicspath{{./Figures/}}
\usepackage[abbreviations, automake]{glossaries-extra}
\makeglossaries 
\newabbreviation{OMA}{OMA}{orthogonal multiple access}
\newabbreviation{NOMA}{NOMA}{Non-orthogonal multiple access}
\newabbreviation{IRS}{IRS}{intelligent reflecting surface}
\newabbreviation{BS}{BS}{base station}
\newabbreviation{LoS}{LoS}{Line-of-sight}
\newabbreviation{AoA}{AoA}{angle of arrival}
\newabbreviation{AoD}{AoD}{angle of departure}
\newabbreviation{SINR}{SINR}{signal-to-interference-plus-noise ratio}
\newabbreviation{ASR}{ASR}{achievable sum rate}
\newabbreviation{PDF}{PDF}{probability distribution function}
\newabbreviation{CDF}{CDF}{cumulative distribution function}
\newabbreviation{5G}{5G}{fifth-generation}
\newabbreviation{MPA}{MPA}{maximum ASR achieving power allocation}
\newabbreviation{FPA}{FPA}{fair power allocation}
\newabbreviation{AUP}{AUP}{adaptive user pairing}
\newabbreviation{EE}{EE}{energy efficient}

\newacronym{M}{\ensuremath{M}}{Number of antennae at the base station}
\newacronym{N}{\ensuremath{N}}{Number of antennae at the IRS}
\newacronym{HI}{\ensuremath{\bold{h}_R}}{Channel between BS to IRS}
\newacronym{hi}{\ensuremath{\bold{h}_i^H}}{Channel between IRS and user $i$}
\newacronym{h}{\ensuremath{h}}{Number of antennae at the IRS}
\newacronym{PLI}{\ensuremath{\beta_I}}{Signal strength of the link between BS to IRS}
\newacronym{PLi}{\ensuremath{\beta_i}}{Signal strength of the link between IRS to the user $i$}
\newacronym{h1}{\ensuremath{\bold{h}_1^H}}{Channel between IRS and user $1$}
\newacronym{PL1}{\ensuremath{\beta_1}}{Signal strength of the link between IRS to the user $1$}
\newacronym{h2}{\ensuremath{\bold{h}_2^H}}{Channel between IRS and user $2$}
\newacronym{PL2}{\ensuremath{\beta_2}}{Signal strength of the link between IRS to the user $2$}
\newacronym{ArrIa}{\ensuremath{\phi_I^a}}{Azimuth angle of arrival at the IRS}
\newacronym{DepIa}{\ensuremath{\psi_I^a}}{Azimuth angle of departure from the IRS}
\newacronym{ArrIe}{\ensuremath{\phi_I^e}}{Zenith angle of arrival at the IRS}
\newacronym{DepIe}{\ensuremath{\psi_I^e}}{Zenith angle of departure from the IRS}
\newacronym{DepBa}{\ensuremath{\psi_B^a}}{Azimuth angle of departure from the BS}
\newacronym{DepBe}{\ensuremath{\psi_B^e}}{Zenith angle of departure from the BS}
\newacronym{a1}{\ensuremath{\alpha_1}}{Power allocated to strong user}
\newacronym{a1F}{\ensuremath{\alpha_1^\text{\tiny FPA}}}{Power allocated to strong user}
\newacronym{a1M}{\ensuremath{\alpha_1^\text{\tiny MPA}}}{Power allocated to strong user}
\newacronym{a2}{\ensuremath{\alpha_2}}{Power allocated to weak user}
\newacronym{Pt}{\ensuremath{P_t}}{Total available transmit power at the BS}
\newacronym{DiagI}{\ensuremath{\bold{\Theta}}}{IRS reflection diagonal matrix}
\newacronym{DiagI'}{\ensuremath{\bold{\widetilde{\Theta}}}}{IRS reflection diagonal matrix with imperfect phase compensation}
\newacronym{delta}{\ensuremath{\delta}}{Maximum possible phase noise}
\newacronym{GammaW}{\ensuremath{\gamma_2^\text{\tiny NOMA}}}{SINR of weak user in a NOMA system}
\newacronym{GammaS}{\ensuremath{\gamma_1^\text{\tiny NOMA}}}{SINR of strong user in a NOMA system}
\newacronym{RateW}{\ensuremath{R_2^\text{\tiny NOMA}}}{SINR of weak user in a NOMA system}
\newacronym{RateS}{\ensuremath{R_1^\text{\tiny NOMA}}}{SINR of strong user in a NOMA system}
\newacronym{RateNoma}{\ensuremath{R_i^\text{\tiny NOMA}}}{SINR of weak user in a NOMA system}
\newacronym{RateO}{\ensuremath{R_i^\text{\tiny OMA}}}{SINR of weak user in a NOMA system}
\newacronym{RateO1}{\ensuremath{\overline{R}_1^\text{\scriptsize }}}{SINR of weak user in a NOMA system}
\newacronym{RateO2}{\ensuremath{\overline{R}_2^\text{\scriptsize }}}{SINR of weak user in a NOMA system}
\newacronym{RateOi}{\ensuremath{\overline{R}_i^\text{\scriptsize }}}{SINR of weak user in a NOMA system}
\newacronym{sigma2}{\ensuremath{\sigma^2}}{Noise power}
\newacronym{I}{\ensuremath{I}}{Interference}
\newacronym{GammaCSI}{\ensuremath{\gamma_i^{\text{\tiny CSI}}}}{SINR of strong user in a OMA system}
\newacronym{Gamma1CSI}{\ensuremath{\gamma_1^\text{\tiny CSI}}}{SINR of weak user in a OMA system}
\newacronym{Gamma2CSI}{\ensuremath{\gamma_2^\text{\tiny CSI}}}{SINR of weak user in a OMA system}
\newacronym{GammaIO}{\ensuremath{\gamma_i^\text{\tiny OMA}}}{SINR of $i^{th}$ user  in a OMA system}
\newacronym{Ow}{\ensuremath{\mathcal{O}_2^\text{\tiny }}}{SINR of $i^{th}$ user  in a OMA system}
\newacronym{Os}{\ensuremath{\mathcal{O}_1^\text{\tiny }}}{SINR of $i^{th}$ user  in a OMA system}
\newacronym{Oi}{\ensuremath{\mathcal{O}_i^\text{\tiny }}}{SINR of $i^{th}$ user  in a OMA system}
\newacronym{thetak'}{\ensuremath{\hat{\theta}_n}}{Maximum possible phase noise}
\DeclareMathOperator{\sinc}{sinc}

\begin{document}
\bstctlcite{IEEEexample:BSTcontrol}
\title{\huge User Pairing and Power Allocation for IRS-Assisted NOMA Systems with Imperfect Phase Compensation}
\author{\IEEEauthorblockN{Pavan Reddy M., Abhinav Kumar}\thanks{Pavan Reddy M. and Abhinav Kumar are with the Department of Electrical Engineering, Indian Institute of Technology Hyderabad, Telangana, India.  ~~~~{(e-mail:\{ee14resch11005,~abhinavkumar\}@iith.ac.in).}
}}
\maketitle
\begin{abstract}
In this letter, we analyze the performance of the intelligent reflecting surface (IRS) assisted downlink non-orthogonal multiple access (NOMA) systems in the presence of imperfect phase compensation. We derive an upper bound on the imperfect phase compensation to achieve minimum required data rates for each user. Using this bound, we propose an adaptive user pairing algorithm to maximize the network throughput. We then derive bounds on the power allocation factors and propose power allocation algorithms for the paired users to achieve the maximum sum rate or ensure fairness.  Through extensive simulations, we show that the proposed algorithms significantly outperform the state-of-the-art algorithms. 
\end{abstract}
\section{Introduction}
\gls{NOMA} is considered as a key radio access technique for \gls{5G} and beyond \gls{5G} networks~\cite{Noma}. In \gls{NOMA}, the users are allocated the same time and frequency resources but are multiplexed across the power domain to achieve multi-fold improvement in the network capacity. At the receiver side of \gls{NOMA} systems, the successive interference cancellation is employed to decode the transmitted data.  Similar to \gls{NOMA}, \gls{IRS} is another key technology to improve coverage for the beyond-\gls{5G} networks~\cite{Coverage}. An \gls{IRS} consists of a large number of passive antenna elements where the reflection from each antenna is controlled to direct the signal towards a particular user.  Motivated by the benefits from both technologies, \gls{IRS} has been analyzed along with \gls{NOMA} to achieve better network capacity and enhanced coverage~\cite{Simple}.  

The practical \gls{IRS} systems have imperfections in the phase control because of hardware limitations and channel estimations errors~\cite{MISO}.  These imperfections in the phase compensation have a significant impact on the data rates observed by the users. However, only a few works in the literature consider these imperfections while analyzing the network performance~\cite{Simple, MISO}. In \cite{Simple}, the authors have proposed a novel design for \gls{IRS}-assisted \gls{NOMA} transmissions and have analyzed the impact of hardware impairments.  In~\cite{MISO}, the authors have evaluated the performance of \gls{OMA} systems in the presence of imperfect phase compensation.  In \gls{NOMA}, the network capacity is heavily dependent on the user pairing~\cite{Mouni, DynamicUsers, GenPairing}, and hence, \gls{IRS}-assisted \gls{NOMA} systems have to consider the imperfections in the phase compensation while pairing the users. Otherwise, the enhanced network throughputs will not be realized in practice. To the best of our knowledge, none of the existing works in the literature have proposed user pairing and power allocation for \gls{IRS}-assisted \gls{NOMA} systems with imperfect phase compensation. 

In view of the aforementioned details, this letter presents the first work that discusses the following contributions.
\begin{itemize}
    \item We derive bounds on the imperfection phase compensation to achieve minimum required data rates.
    \item We propose an adaptive user pairing algorithm for \gls{IRS}-assisted downlink \gls{NOMA} systems.
    \item We derive various bounds on the power allocation factors for the paired users.
    \item Using the derived bounds, we propose power allocation algorithms to maximize the \gls{ASR} or ensure fairness. 
\end{itemize}
The organization of the paper is as follows. We present the system model in Section~\ref{sec:SysModel}. In Section~\ref{sec:CoB}, we derive various bounds on the imperfect phase compensation and power allocation factors. We propose  adaptive user pairing and power allocation algorithms in Section~\ref{sec:Proposed}. In Section~\ref{sec:Results}, we present numerical results for various scenarios. We then provide concluding remarks and directions for future work in Section~\ref{sec:Conclusion}.

\section{System Model}
\label{sec:SysModel}
\begin{figure}
    \centering
    \includegraphics[scale=0.53]{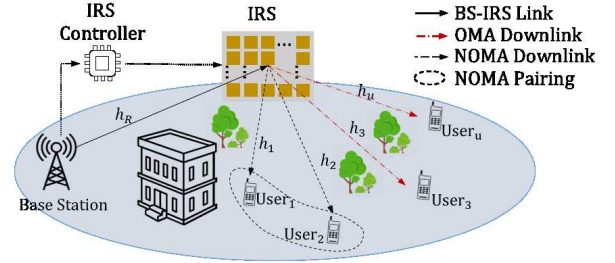}
    \caption{System Model.}
    \label{fig:SysModel}
\end{figure}
We consider a \gls{BS} with \gls{M} antennae and an \gls{IRS} with \gls{N} antennae, where \gls{IRS} is activated by a controller connected to the \gls{BS} as shown in Fig.~\ref{fig:SysModel}. 
The channel coefficients between the \gls{BS} to \gls{IRS} and  $i^{th}$ user to \gls{IRS} are denoted by \gls{HI} and $\bold{h}_i$, respectively, and are defined as follows~\cite{MISO}:
\begin{align}
    \gls{HI}=&\gls{PLI}   \bold{a}_N(\gls{ArrIa},\gls{ArrIe}) \bold{a}_M^H(\gls{DepBa},\gls{DepBe}),\label{eqn:HI}\\
    \bold{h}_i=& \gls{PLi}   \bold{a}_N(\gls{DepIa},\gls{DepIe}),\label{eqn:hi}
\end{align}
where $\{\cdot\}^H$ is the Hermitian of the matrix, \gls{PLI} and \gls{PLi} are the distance dependent losses of \gls{BS} to \gls{IRS} link and \gls{IRS} to $i^{th}$ user link, respectively, \gls{ArrIa} and \gls{ArrIe} are the \gls{AoA} in azimuth and elevation at the \gls{IRS}, respectively, \gls{DepBa} and \gls{DepBe} are the \gls{AoD} in azimuth and elevation at the \gls{BS}, respectively, \gls{DepIa} and \gls{DepIe} are the \gls{AoD} in the azimuth and elevation at the \gls{IRS}, respectively, and $\bold{a}_X(\upsilon^a,\upsilon^e)$ is the array factor that captures the beamforming gain. For a planar array with $X$ antenna elements, we assume $\sqrt{X}$ elements in the horizontal and vertical direction of the planar array, and thus, define the array factor as follows~\cite{MISO}:
\vspace{-0.2cm}
{ \begin{align}
  \bold{a}_X(\upsilon^a,\upsilon^e)&=
  \begin{bmatrix} 
  1\\\vdots\\e^{j\frac{2\pi d}{\lambda}(x\sin\upsilon^a\sin\upsilon^e+y\cos\upsilon^e)}\\\vdots\\e^{j\frac{2\pi d}{\lambda}\big((\sqrt X -1)\sin\upsilon^a\sin\upsilon^e+(\sqrt X-1)\cos\upsilon^e\big)}
  \end{bmatrix}
\nonumber 
\end{align}
{\normalsize where $0\leq x,y\leq(\sqrt X-1)$ are the indices of antenna elements in the planar array,  $d$ is the spacing between antenna elements, $\lambda$ is the wavelength, $\upsilon^a$ and $\upsilon^e$ are the desired directions in azimuth and elevation, respectively.}}

We denote the diagonal matrix that captures the reflection of the \gls{IRS} as \gls{DiagI} and define each diagonal element of \gls{DiagI} as $e^{j\theta_k}$~\cite{Simple}, where $k\in[1,\gls{N}]$ is the antenna index and $\theta_k~\in~[0,2\pi)$ is the phase reflection coefficient. Further, practical \gls{IRS} have finite resolution while applying phase shifters, and hence, result in imperfect phase compensation. Thus, we consider the actual reflection matrix to be $\gls{DiagI'}$ with each diagonal element defined as $e^{j\tilde{\theta}_k}$, where $\tilde{\theta}_k=\theta_k+\hat{\theta}_k$, $\hat{\theta}_k$ being the phase noise. We consider $\hat{\theta}_k$ to be uniformly distributed over $[-\gls{delta},\gls{delta}]$ with $\gls{delta}\in[0,\pi)$. With all this information, the signal received by the $i^{th}$ user in an \gls{OMA} is formulated as~\cite{Simple}
\begin{align}
    y_i^\text{\tiny OMA}=\gls{hi}\gls{DiagI'}\gls{HI}\gls{Pt} s_i+n, \nonumber
\end{align}
where $s_i$ is the data transmitted to $i^{th}$ user, $n$ denotes the noise,  and $\gls{Pt}$ is the available transmit power at the \gls{BS}. The \gls{SINR} of the $i^{th}$ user in the \gls{OMA} system is formulated as
\begin{align}
\gls{GammaIO}=&\dfrac{\gls{Pt}||\gls{hi}\gls{DiagI'}\gls{HI}||^2}{\gls{I}+\gls{sigma2}},\label{eqn:GammaO}
\end{align}
where \gls{sigma2} is the noise variance and \gls{I} is the interference power received at the user.
In case of an \gls{IRS}-assisted downlink \gls{NOMA} system, we consider that the \gls{BS} transmits $\gls{Pt}(\gls{a1} s_1+\gls{a2}s_2)$, where $\gls{a1}$ and $\gls{a2}$ are the fractions of power allocated to strong and weak user, respectively, and $s_1$ and $s_2$ denote the data to be transmitted to the strong and weak user, respectively. Further, $0<\gls{a1},\gls{a2}<1$ and $\gls{a1}+\gls{a2}=1$. The signal received by the $i^{th}$ user in \gls{NOMA} is formulated as~\cite{Simple}
\begin{align}
    y_i^\text{\tiny NOMA}=\gls{hi}\gls{DiagI'}\gls{HI}\gls{Pt}(\gls{a1}s_2+\gls{a2}s_2)+n, \forall i\in 1, 2. \nonumber
\end{align}
Thus, we define the \gls{SINR} of strong and weak users in a \gls{NOMA} system as follows:
\begin{align}
    \gls{GammaS}&=\dfrac{\gls{a1}\gls{Pt}||\gls{h1}\gls{DiagI'}\gls{HI}||^2}{\gls{I}+\gls{sigma2}}\label{eqn:Gamma1N},\\
    \gls{GammaW}&=\dfrac{\gls{a2}\gls{Pt}||\gls{h2}\gls{DiagI'}\gls{HI}||^2}{{\gls{a1}\gls{Pt}||h_2\gls{DiagI'}\gls{HI}||^2}+\gls{I}+\gls{sigma2}}.\label{eqn:Gamma2N}
\end{align}
\section{Computation of Bounds}
\label{sec:CoB}
We derive the achievable data rates by users in \gls{IRS}-assited \gls{NOMA} and \gls{OMA} systems as follows.~\\From \eqref{eqn:HI}-\eqref{eqn:hi}, we get
\begin{align}
\gls{hi}\gls{DiagI'}\gls{HI}&=\gls{PLi}\gls{PLI}\sum_{n=1}^{\gls{N}} e^{j\gls{thetak'}}\bold{a}_M^H(\gls{DepBa},\gls{DepBe}),\label{eqn:c1}\\
|| \bold{a}_M^H(\gls{DepBa},\gls{DepBe})||^2&=M,\label{eqn:c2}\\
||\gls{hi}\gls{DiagI'}\gls{HI}||^2&=|\gls{PLi}\gls{PLI}|^2\Big|\sum_{n=1}^{\gls{N}} e^{j\gls{thetak'}}\Big|^2 M. \label{eqn:c3}
\end{align}
We define channel state information (CSI) of $i^{th}$ user (\gls{GammaCSI}) as
\begin{align}
\gls{GammaCSI}=&\dfrac{\gls{Pt}||\gls{hi}\gls{DiagI}\gls{HI}||^2}{\gls{I}+\gls{sigma2}}
=\dfrac{\gls{Pt}|\gls{PLi}\gls{PLI}|^2\big|N^2 M}{\gls{I}+\gls{sigma2}}.\label{eqn:GammaCSI}
\end{align}
\begin{lemma}
The normalized achievable data rates in an IRS-assisted \gls{OMA} and \gls{NOMA} systems with 2 users are as follows:
\begin{align}
\gls{RateO}=&\dfrac{1}{2}\log_2\big(1+\gls{GammaCSI}\sinc^2(\gls{delta})\big), \forall i \in 1,2,\label{eqn:R0}\\
\gls{RateS}=&\log_2\big(1+\gls{a1}\gls{Gamma1CSI}\sinc^2(\gls{delta})\big),\label{eqn:R1}\\
\gls{RateW}=&\log_2\Big(1+\dfrac{\gls{a2}\gls{Gamma2CSI}\sinc^2(\gls{delta})}{\gls{a1}\gls{Gamma2CSI}\sinc^2(\gls{delta})+1}\Big)\label{eqn:R2}.
\end{align}
\end{lemma}

\begin{proof}
We adopt the \gls{SINR} approximation formulated in~\cite{MISO} and define the following:
\begin{align}
    \Bigg|\dfrac{1}{N}\sum_{n=1}^{\gls{N}} e^{j\gls{thetak'}}\Bigg|^2 \overset{\text{\scriptsize(a)}}{\longrightarrow} \big|\mathbb{E}[e^{j\gls{thetak'}}]\big|^2\overset{\text{\scriptsize (b)}}{=} \big|\mathbb{E}[\cos(\gls{thetak'})]\big|^2\overset{\text{\scriptsize (c)}}{=}\sinc^2(\gls{delta}),
\label{eqn:sinc2}
\end{align}
where $(a)$ is based on law of large numbers~\cite{MISO}, $(b)$ is obtained by integrating the odd symmetrical function  $\sin(\gls{thetak'})$ for $\gls{thetak'}\in[-\gls{delta}, \gls{delta}]$, and $(c)$ uses the probability density function of \gls{thetak'} which is defined as $f(\gls{thetak'})={1}/{2\gls{delta}}, \forall \gls{thetak'}\in[-\gls{delta},\gls{delta}]$ and ${\sinc (\gls{delta})={\sin (\gls{delta})}/{\gls{delta}}}$. Substituting \eqref{eqn:c1}-\eqref{eqn:GammaCSI} and \eqref{eqn:sinc2} in \eqref{eqn:GammaO}-\eqref{eqn:Gamma2N}, we get
\begin{align}
\gls{GammaIO}=&\dfrac{\gls{Pt}|\gls{PLi}\gls{PLI}|^2\big|\sum_{n=1}^{\gls{N}} e^{j\gls{thetak'}}\big|^2 M}{\gls{I}+\gls{sigma2}}=\gls{GammaCSI}\sinc^2(\gls{delta}), \forall i \in 1,2,\label{eqn:GammaOU}\\
\gls{GammaS}=&\dfrac{\gls{a1}\gls{Pt}|\gls{PL1}\gls{PLI}|^2\big|\sum_{n=1}^{\gls{N}} e^{j\gls{thetak'}}\big|^2 M}{\gls{I}+\gls{sigma2}}=\gls{a1}\gls{Gamma1CSI}\sinc^2(\gls{delta}),\label{eqn:Gamma2NU}\\
\gls{GammaW}=&\dfrac{\gls{a2}\gls{Pt}|\gls{PL2}\gls{PLI}|^2\big|\sum_{n=1}^{\gls{N}} e^{j\gls{thetak'}}\big|^2 M}{\gls{a1}\gls{Pt}|\gls{PL2}\gls{PLI}|^2\big|\sum_{n=1}^{\gls{N}} e^{j\gls{thetak'}}\big|^2 M+\gls{I}+\gls{sigma2}},\nonumber\\
=&\dfrac{\gls{a2}\gls{Gamma2CSI}\sinc^2(\gls{delta})}{\gls{a1}\gls{Gamma2CSI}\sinc^2(\gls{delta})+1}.
\label{eqn:Gamma2NU}
\end{align}
Assuming the full bandwidth allocation for the two users in case of \gls{NOMA} and half bandwidth allocation for each user in \gls{OMA}, and substituting \eqref{eqn:GammaOU}-\eqref{eqn:Gamma2NU} while calculating the normalized data rates, we complete the proof of Lemma~1. 
\end{proof}
Next, we derive bounds on the power allocation factors.
\subsection{Bounds on \gls{a1} and \gls{a2}}
We define \gls{RateO1} and \gls{RateO2} as the minimum rates required by the strong and the weak user, respectively.
For the lower bound on \gls{a1}, we assume that rate of weak user in \gls{NOMA} (\gls{RateS}) should be greater than or equal to the minimum rate required by the weak user (\gls{RateO1}). Thus, by considering $\gls{RateS}\geq \gls{RateO1}$, we get
\begin{align}
\log_2\big(1+\gls{a1}\gls{Gamma1CSI}\sinc^2(\gls{delta})\big) \geq &\gls{RateO1},\nonumber\\
\gls{a1}\geq &\dfrac{2^{\gls{RateO1}}-1}{\gls{Gamma1CSI}\sinc^2(\gls{delta})}\triangleq \gls{a1}_\text{\tiny LB}.
\label{eqn:LbAlpha}
\end{align}
Similarly, for the upper bound, by using  $\gls{RateW}>\gls{RateO2}$, we get
\begin{align}
\log_2\Big(1+\dfrac{\gls{a2}\gls{Gamma2CSI}\sinc^2(\gls{delta})}{\gls{a1}\gls{Gamma2CSI}\sinc^2(\gls{delta})+1}\Big)\geq&\gls{RateO2}.
\label{eqn:Rate2}
\end{align}
Substituting $\gls{a2}=1-\gls{a1}$ in \eqref{eqn:Rate2}, we obtain
\begin{align}
\gls{a1}\leq\dfrac{\gls{Gamma2CSI}\sinc^2(\gls{delta})-(2^{\gls{RateO2}}-1)}{2^{\gls{RateO2}}\gls{Gamma2CSI}\sinc^2(\gls{delta})}\triangleq \gls{a1}_\text{\tiny UB}.
\label{eqn:UbAlpha}
\end{align}
Note that similar bounds can be achieved for the power allocation factor of the weak user by substituting $\gls{a2}=1-\gls{a1}$ in \eqref{eqn:LbAlpha}-\eqref{eqn:UbAlpha}.
\subsection{Upper Bound on the Imperfect Phase Compensation ($\gls{delta}_\text{\tiny UB}$)}
For the upper bound on \gls{delta}, we consider that the upper bound of \gls{a1} in \eqref{eqn:UbAlpha} should be greater than or equal to the lower bound of \gls{a1} in \eqref{eqn:LbAlpha}. Using \eqref{eqn:LbAlpha}-\eqref{eqn:UbAlpha} and solving $\gls{a1}_\text{\tiny UB}~\geq~\gls{a1}_\text{\tiny LB}$, we get
\begin{align}
\sinc^2(\gls{delta})\geq\dfrac{(2^{\gls{RateO1}}-1)2^{\gls{RateO2}}}{\gls{Gamma1CSI}}+\dfrac{(2^{\gls{RateO2}}-1)}{\gls{Gamma2CSI}}\triangleq \sinc^2(\gls{delta}_\text{\tiny UB}).
\label{eqn:UbSinc}
\end{align}
Note that when we consider $\overline{R}_i=\gls{RateO}$, $\gls{delta}_\text{\tiny UB}$ is computable at the base station as it is only dependent on \gls{GammaCSI}. From~\eqref{eqn:UbSinc}, we conclude that it is beneficial to pair the users in \gls{IRS}-assisted \gls{NOMA} systems only when $\gls{delta}_\text{\tiny UB}$ with that user pair  is greater than or equal to \gls{delta}. Otherwise, the data rates achieved by the users in \gls{NOMA}  will not be higher than their \gls{OMA} counterparts. 

\section{Proposed Algorithms}
\label{sec:Proposed}
In this section, we initially present an \gls{AUP} algorithm for the \gls{IRS}-assisted \gls{NOMA} system based on the $\gls{delta}_\text{\tiny UB}$ derived in \eqref{eqn:UbSinc}. We then propose \gls{MPA} and \gls{FPA} algorithms for the paired users.
\subsection{\gls{AUP}}
\begin{figure}
    \centering
    \includegraphics[scale=0.5]{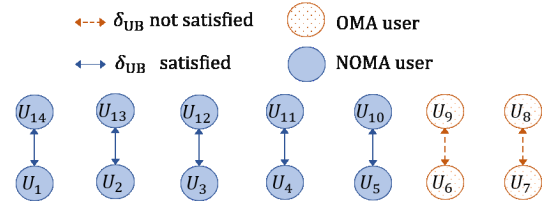}
    \caption{Adaptive user pairing.}
    \label{fig:IRSAUP}
\end{figure}
In~\cite{NearFar1,NearFar2,GenPairing}, the authors have shown that pairing of near users with far users achieves better data rates. Motivated by this, we initially sort the users based on their \gls{SINR}s  and group the near users with far users. For each user, we then define the rate achievable with \gls{OMA} as the minimum required rate (i.e., $\overline{R_i}~=~\gls{RateO}$). From \eqref{eqn:UbSinc}, it is evident that pairing two users in \gls{IRS}-assisted \gls{NOMA} with imperfect phase compensation will not always ensure that achievable data rates are better than \gls{OMA} rates. Hence, to exploit the benefits from \gls{NOMA}, we pair only those users whose achievable date rates outperform the \gls{OMA} counterparts. Thus, for users in each group, we check if the imperfect phase compensation (\gls{delta}) is less than or equal to  $\gls{delta}_\text{\tiny UB}$ formulated in \eqref{eqn:UbSinc}. If this criterion is satisfied, we consider the users in that group to be a \gls{NOMA} pair. Otherwise, we consider them to be \gls{OMA} users. This procedure will ensure that each user achieves at least \gls{OMA} rates. 
All this procedure is pictorially presented in Fig.~\ref{fig:IRSAUP} for a set of 14 users with $\gls{Gamma1CSI}\geq...\geq\gamma_{14}^\text{\tiny CSI}$. Next, we present \gls{MPA} procedure in detail.
\vspace{-0.2cm}
\subsection{\gls{MPA}}
In this section, we present maximum \gls{ASR} achieving power allocation procedure for the \gls{IRS}-assisted \gls{NOMA} systems.
\begin{lemma}
\label{lemma:2} 
The power allocation factors for \gls{NOMA} pair that maximize the \gls{ASR} and also ensure each user achieves at least \gls{OMA} rates are as follows, $\gls{a1}=\gls{a1}_\text{\tiny UB}$ and $\gls{a2}=1-\gls{a1}$.
\end{lemma}
\begin{proof}
We formulate \gls{ASR} for a \gls{NOMA} pair as $\gls{RateS}+\gls{RateW}$. 
\begin{align}
    \frac{d(\gls{ASR})}{d \gls{a1}}=\dfrac{\gls{Gamma1CSI}\sinc^2(\gls{delta})-\gls{Gamma2CSI}\sinc^2(\gls{delta})}{(1+\gls{a1}\gls{Gamma1CSI}\sinc^2(\gls{delta}))(1+\gls{a2}\gls{Gamma2CSI}\sinc^2(\gls{delta}))}.\label{eqn:Monoton}
\end{align}Note that as per our formulation in~\eqref{eqn:Gamma1N}-\eqref{eqn:Gamma2N}, $\gls{Gamma1CSI}\geq\gls{Gamma2CSI}$, and thus, $\frac{d(\gls{ASR})}{d\gls{a1}}\geq0$. Hence, \gls{ASR} is a non-decreasing function and $\gls{a1}=\gls{a1}_\text{\tiny UB}$ will result in maximum \gls{ASR}. This completes the proof of Lemma~\ref{lemma:2}.
\end{proof}
Thus, in \gls{MPA}, we allocate $\gls{a1}=\gls{a1}_\text{\tiny UB}$ and $\gls{a2}=1-\gls{a1}$ to strong and weak users, respectively, to achieve maximum~\gls{ASR}.
\subsection{\gls{FPA}}
We define the \gls{Oi} as an event of outage for $i^{th}$ user, where $Pr(\gls{Oi})=Pr(\gls{RateNoma}<\gls{RateOi}), \forall i=1,2.$ Using~\eqref{eqn:R1}, we get \begin{align}
    Pr(\gls{Os})&=Pr(\gls{RateS}<\gls{RateO1}),\nonumber\\
    &=Pr\Big(\gls{Gamma1CSI}<\dfrac{(2^{\gls{RateO1}}-1)}{\gls{a1}\sinc^2(\gls{delta})}\Big).
\label{eqn:Pr1}
\end{align}
Similarly, using \eqref{eqn:R2} for the weak user in \gls{NOMA}, we get
\begin{align}
    Pr(\gls{Ow})&=Pr(\gls{RateW}<\gls{RateO2}),\nonumber\\
    &=Pr\Big(\gls{Gamma2CSI}<\dfrac{(2^{\gls{RateO2}}-1)}{\big(\gls{a2}-\gls{a1}(2^{\gls{RateO2}}-1)\big)\sinc^2(\gls{delta})}\Big). 
\label{eqn:Pr2}
\end{align}
In \gls{NOMA}, with an increase in power allocation for a user, the probability of outage increases for the other paired user. Hence, to ensure fairness in power allocation, we consider allocating power levels such that the probability of outage is same for both the users. Thus, for a given probability distribution function for \gls{GammaCSI}, using \eqref{eqn:Pr1}-\eqref{eqn:Pr2}, when the probability of outage is same for both the paired users, the following holds:
\begin{align}
    \dfrac{(2^{\gls{RateO1}}-1)}{\gls{a1}\sinc^2(\gls{delta})}=\dfrac{(2^{\gls{RateO2}}-1)}{\big(\gls{a2}-\gls{a1}(2^{\gls{RateO2}}-1)\big)\sinc^2(\gls{delta})} .
\label{eqn:Pr1aw}
\end{align}
Substituting $\gls{a2}=1-\gls{a1}$ and solving it further, we obtain\begin{align}
    \gls{a1}=\dfrac{(2^{\gls{RateO1}}-1)}{\big( 2^{\gls{RateO2}}-1+2^{\gls{RateO1}} (2^{\gls{RateO1}}-1)\big)}. 
\label{eqn:Pr1aw}
\end{align}
Note that, we also obtain \eqref{eqn:Pr1aw} by considering the upper bound in \eqref{eqn:UbAlpha} equal to the lower bound in \eqref{eqn:Pr1aw}. Further, \gls{a1} obtained in \eqref{eqn:LbAlpha} satisfies $0<\gls{a1}<1$. A pseud-code to implement the proposed \gls{AUP}, \gls{MPA}, and \gls{FPA} is presented in Algorithm~\ref{Algo:proposed}. The \gls{AUP} algorithm with \gls{MPA} and \gls{FPA} requires sorting of the users, and hence, the complexity is of the order $\mathcal{O}(G\log_2 G)$. Next, we present the simulation results. 
\SetAlCapNameFnt{\small}
\SetAlCapFnt{\small}
\begin{algorithm}[t]  
\small
\setstretch{0.9}
\SetKwInOut{Input}{Input}
\SetKwInOut{Var}{Variables}
\Input{Set of users $U_i$ and corresponding \gls{SINR}s $\gls{GammaCSI}$, imperfection in phase compensation \gls{delta} at the \gls{IRS}.}
\Var{$i$ is a variable representing user pair index.}
\BlankLine
\caption{Proposed \gls{AUP}, \gls{MPA}, and \gls{FPA}}
\label{Algo:proposed} 
Sort the users based on their \gls{SINR}s\ ($\gls{Gamma1CSI}\geq\ldots\geq \gamma_G^\text{\tiny CSI})$\;
\For{$i=1 \rightarrow \frac{G}{2}$}
{$\gls{RateO1}=\frac{1}{2} \log_2(1+\gls{GammaCSI}\sinc^2(\gls{delta}))$\;
$\gls{RateO2}=\frac{1}{2} \log_2(1+\gamma_{G-i+1}^\text{\tiny CSI}\sinc^2(\gls{delta}))$\;
$\sinc^2(\gls{delta}_\text{\tiny \it UB})=\dfrac{(2^{\gls{RateO1}}-1)2^{\gls{RateO2}}}{\gls{Gamma1CSI}}+\dfrac{(2^{\gls{RateO2}}-1)}{\gamma_{G-i+1}^\text{\tiny CSI}}$\;
\eIf{$\sinc^2(\gls{delta}_\text{\tiny UB})>\sinc^2(\gls{delta})$}
{Consider the users for \gls{OMA};}
{Consider the users for \gls{NOMA}\;
\eIf {\gls{MPA}}
{$\gls{a1}=\dfrac{\gamma_{G-i+1}^\text{\tiny CSI}\sinc^2(\gls{delta})-(2^{\gls{RateO2}}-1)}{2^{\gls{RateO2}}\gamma_{G-i+1}^\text{\tiny CSI}\sinc^2(\gls{delta})}
$\;}
{\textit{\gls{FPA}}\;
  $\gls{a1}=\dfrac{(2^{\gls{RateO1}}-1)}{\big( 2^{\gls{RateO2}}-1+2^{\gls{RateO1}} (2^{\gls{RateO1}}-1)\big)}$\;}
  $\gls{a2}=1-\gls{a1}$\;
  }}
\end{algorithm}
\vspace{-0.2cm}
\section{Numerical Results}
\label{sec:Results}
For the evaluation, we have considered Poisson point distributed \gls{BS}s and users with densities 25~BS/km$^2$ and 2000~users/km$^2$, respectively. Further, we have assumed ${\gls{M}=8},\ {N=32},\ {\gls{delta}=11^\circ}, \gls{RateOi}=\gls{RateO}$, and the urban cellular path loss model as presented in \cite{38901}.

\begin{figure}
    \raggedright
    \begin{subfigure}{0.22\textwidth}
    \raggedright
    \includegraphics[scale=0.52]{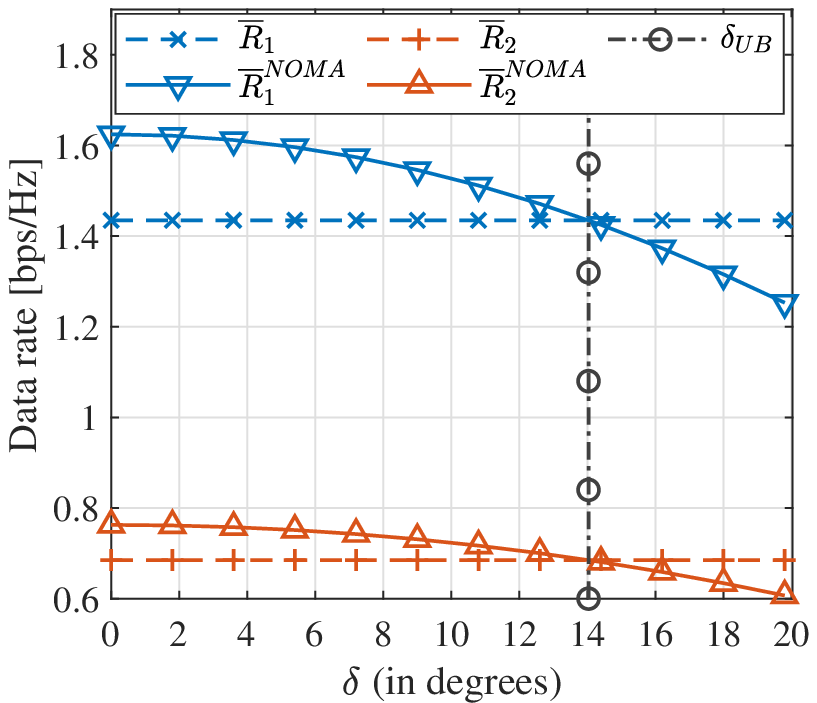}
    \caption{{$[\gls{Gamma1CSI}, \gls{Gamma2CSI}]=[8, 5]~\text{dB}$}.}
    \label{fig:DRa}
    \end{subfigure} \hspace{0.1cm}
    \begin{subfigure}{0.22\textwidth}
    \centering
    \includegraphics[scale=0.52]{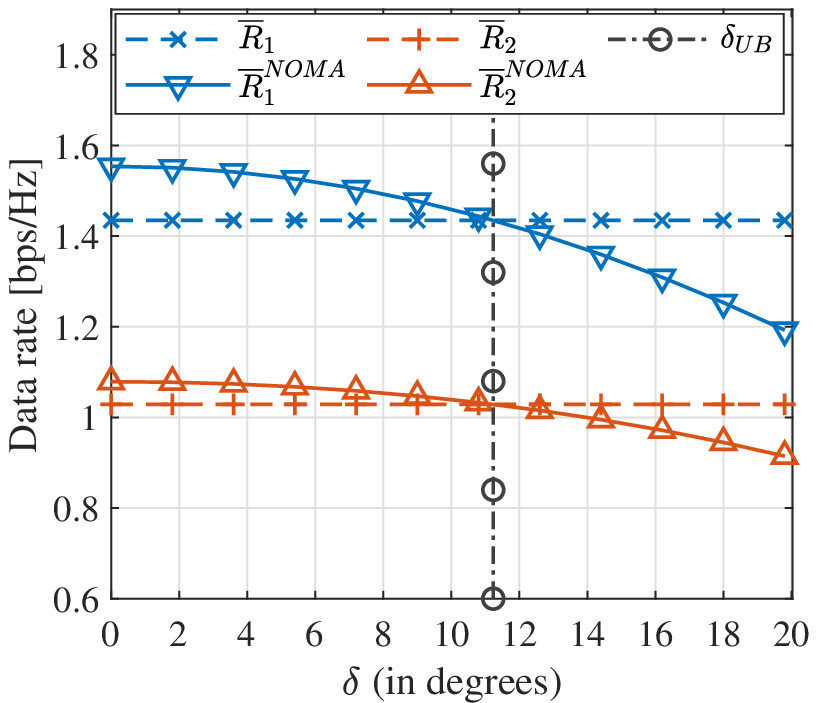}
    \caption{{$[\gls{Gamma1CSI}, \gls{Gamma2CSI}]=[8, 2]~\text{dB} $}.}
    \label{fig:DRb}
    \end{subfigure}
    \caption{Comparison of achievable data rates with varying \gls{delta}.}
    \label{fig:DR}
\end{figure}

In Fig.~\ref{fig:DR}, we present the comparison of data rates of strong and weak users for varying \gls{delta}. For analyzing the impact of \gls{delta}, we have considered a set of \gls{NOMA} user pairs with ${[\gls{Gamma1CSI},\gls{Gamma2CSI}]}=[8, 5]~$dB and $[\gls{Gamma1CSI},\gls{Gamma2CSI}]= [8, 2]~$dB in Fig.~\ref{fig:DRa} and Fig.~\ref{fig:DRb}, respectively. However, note that a similar behaviour holds for any \gls{NOMA} user pair. 
In Fig.~\ref{fig:DRa} and \ref{fig:DRb}, the minimum required rate for strong user (\gls{RateO1}) is same, however, its \gls{NOMA} rates (\gls{RateS}) vary as they depend on the \gls{SINR} of the other paired user.  Further, $\gls{delta}_\text{\tiny UB}$ is different for both the pairs as it is a function of individual \gls{SINR}s of the users in the \gls{NOMA} pair. Also note that, only when $\gls{delta}<\gls{delta}_\text{\tiny UB}$, the \gls{NOMA} rates for both strong and weak users are better than the minimum required rates. Hence, for \gls{NOMA} rates to be better than \gls{OMA} rates, the base station should consider $\gls{delta}_\text{\tiny UB}$ while pairing the users.

\begin{figure*}
    \raggedright
    \begin{subfigure}{0.24\textwidth}
    \raggedright
    \includegraphics[scale=0.5]{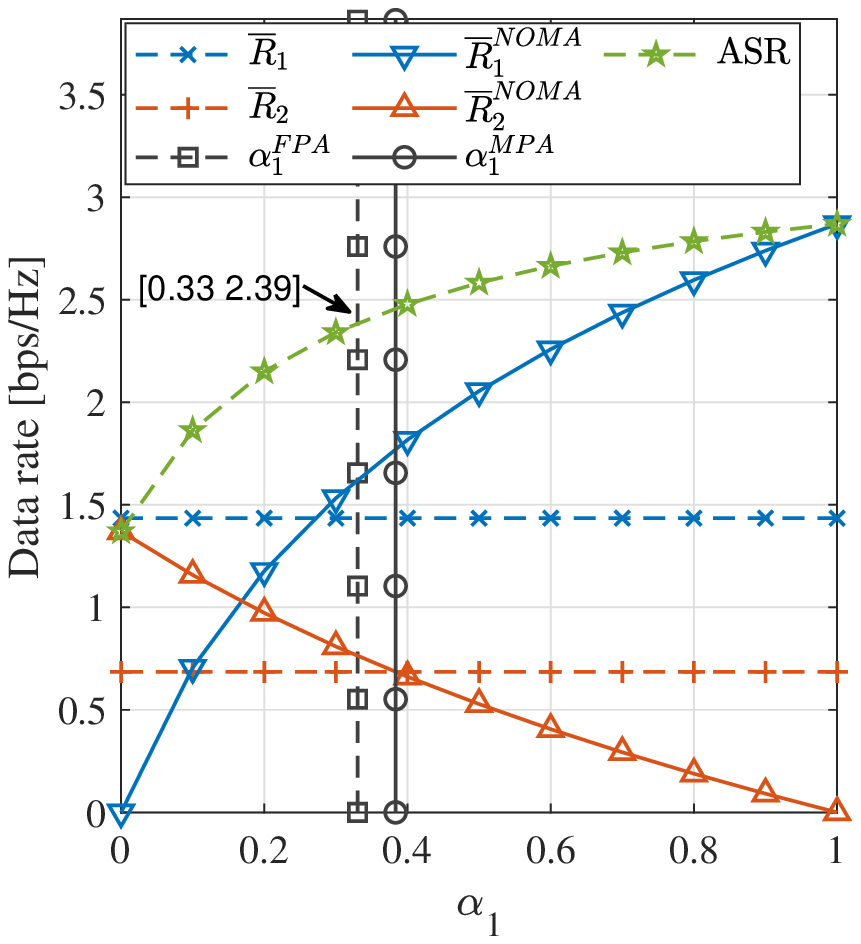}
    \caption{{\footnotesize  $[\gls{Gamma1CSI}, \gls{Gamma2CSI}]=[8, 5]~\text{dB}, \gls{delta}=0^\circ $}.}
    \label{fig:DA1}
    \end{subfigure} \hspace{0.0cm}
    \begin{subfigure}{0.24\textwidth}
    \centering
    \includegraphics[scale=0.5]{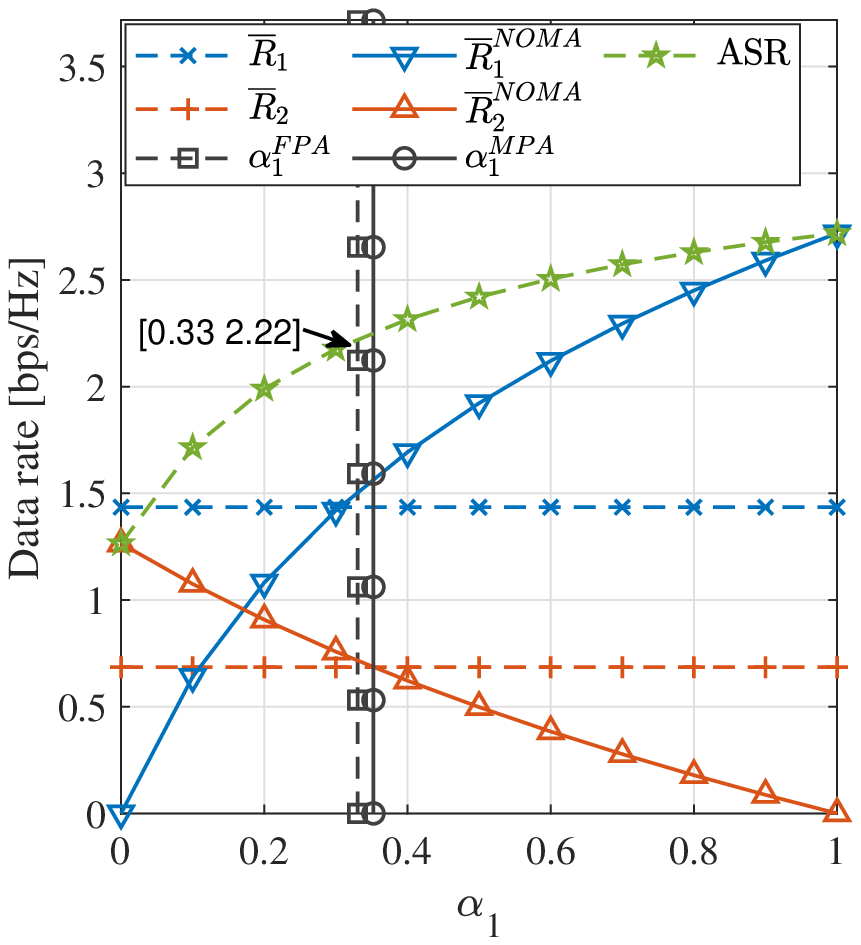}
    \caption{{\footnotesize  $[\gls{Gamma1CSI}, \gls{Gamma2CSI}]=[8, 5]~\text{dB}, \gls{delta}=11^\circ $}.}
    \label{fig:DA2}
    \end{subfigure} \hspace{0cm}    
    \begin{subfigure}{0.24\textwidth}
    \centering
    \includegraphics[scale=0.5]{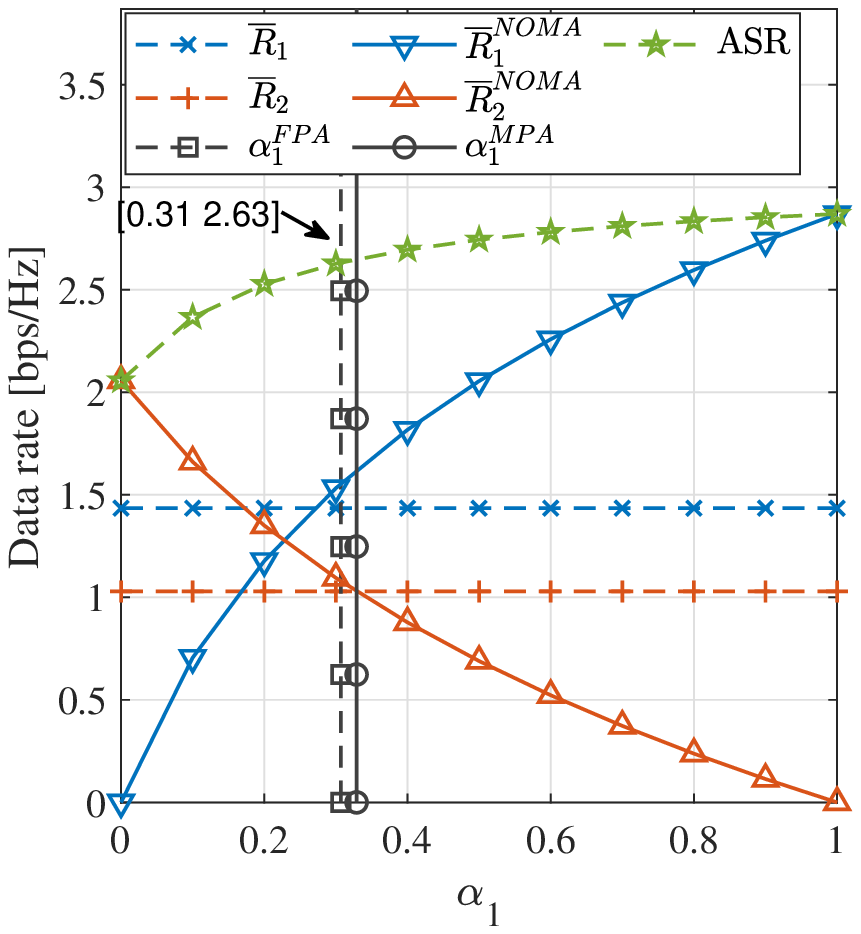}
    \caption{{\footnotesize  $[\gls{Gamma1CSI}, \gls{Gamma2CSI}]=[8, 2]~\text{dB}, \gls{delta}=0^\circ $}.}
    \label{fig:DA3}
    \end{subfigure} \hspace{0cm}
    \begin{subfigure}{0.24\textwidth}
    \centering
    \includegraphics[scale=0.5]{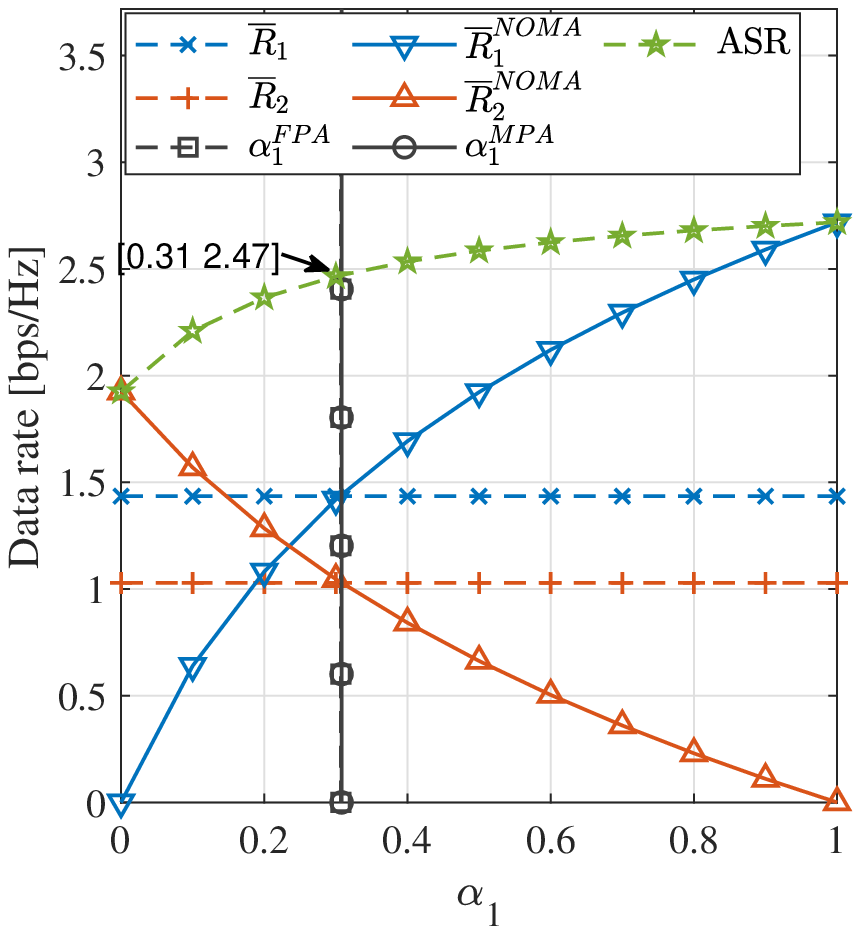}
    \caption{{\footnotesize  $[\gls{Gamma1CSI}, \gls{Gamma2CSI}]=[8, 2]~\text{dB}, \gls{delta}=11^\circ $}.}
    \label{fig:DA4}
    \end{subfigure}    
    \caption{Comparison of achievable data rates with varying \gls{a1}.}    
    \label{fig:DA}
\end{figure*}

In Fig.~\ref{fig:DA}, we present the comparison of data rates with varying $\gls{a1}$. We consider $ {[\gls{Gamma1CSI},\gls{Gamma2CSI}]}=[8, 5]~$dB with $\gls{delta}=0^\circ$ and $11^\circ$ in Fig.~\ref{fig:DA1} and \ref{fig:DA2}, respectively. Since the \gls{SINR}s of the users in \gls{NOMA} pair are same, the minimum required rates by the users (\gls{RateO1} and \gls{RateO2}) are same in both the cases. However, the individual \gls{NOMA} rates vary with \gls{delta} and are better in case of $\gls{delta}=0^\circ$. As shown in Fig.~\ref{fig:DA1} and \ref{fig:DA2}, the power allocation \gls{a1} chosen by \gls{MPA} is the upper bound, beyond which the data rates of weak user will be less than the minimum required rate. Further, observe that \gls{ASR} is a non-decreasing function as presented in \eqref{eqn:Monoton}. Hence, the \gls{ASR} obtained at $\gls{a1M}$ is always higher than $\gls{a1F}$. Additionally, \gls{ASR} is better when $\gls{delta}=0^\circ$  in Fig.~\ref{fig:DA1} as compared to  $\gls{delta}=11^\circ$ in Fig.~\ref{fig:DA2}. In Fig.~\ref{fig:DA3} and \ref{fig:DA4}, we consider $ {[\gls{Gamma1CSI},\gls{Gamma2CSI}]}=[8, 2]~$dB with $\gls{delta}=0^\circ$ and $11^\circ$, respectively. Compared to Fig.~\ref{fig:DA1} and \ref{fig:DA2}, the \gls{ASR} at $\gls{a1F}$ is higher in Fig.~\ref{fig:DA3} and \ref{fig:DA4}. This is because, $\gls{a1F}$ is comparatively less in the latter case, and hence, the interference observed by the weak user is less which results in better \gls{ASR}. Note that with increasing \gls{delta}, the gap between the lower bound in \eqref{eqn:LbAlpha} and the upper bound in \eqref{eqn:UbAlpha} decreases, and thus, the gap between $\gls{a1F}$ and $\gls{a1M}$ also decreases. 
\begin{figure}
    \centering
    \includegraphics[scale=0.48]{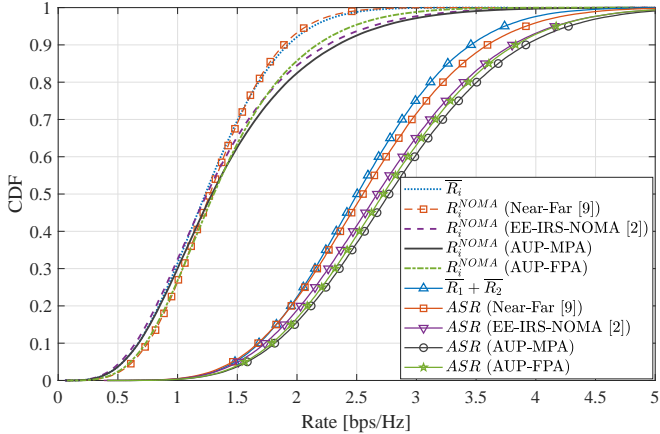}
    \caption{CDF of achievable data rates with various algorithms.}
    \label{fig:cdf}
\end{figure}

In Fig.~\ref{fig:cdf}, we present the \gls{CDF} of achievable data rates with various algorithms. With the Near-Far algorithm~\cite{NearFar2} and \gls{EE} algorithm \gls{EE}-\gls{IRS}-\gls{NOMA}~\cite{Coverage}, the individual \gls{NOMA} rates for some users are worse than the minimum required rates. However, when we consider the proposed \gls{AUP}, the individual rates are never worse than the minimum required rates. Further, when \gls{AUP} is used along with \gls{MPA}, the strong users observe higher data rates and the weak users observe the minimum required data rates, as the algorithm allocates more power to the strong user. Even though Near-Far~\cite{NearFar2} has poor individual rates for some users, it has higher \gls{ASR} as compared to the minimum required \gls{ASR}. The \gls{EE}-\gls{IRS}-\gls{NOMA} algorithm tries to maximize the data rates of the paired \gls{NOMA} users, and thus, has higher \gls{ASR} than the Near-Far algorithm. The proposed \gls{AUP} with \gls{FPA} and \gls{MPA} algorithms have significant improvements in terms of \gls{ASR} as compared to the minimum required \gls{ASR}. Further, \gls{MPA} has comparatively higher \gls{ASR} than \gls{FPA}, as it allocates more power to the strong user.

\begin{figure}
    \centering
    \includegraphics[scale=0.48]{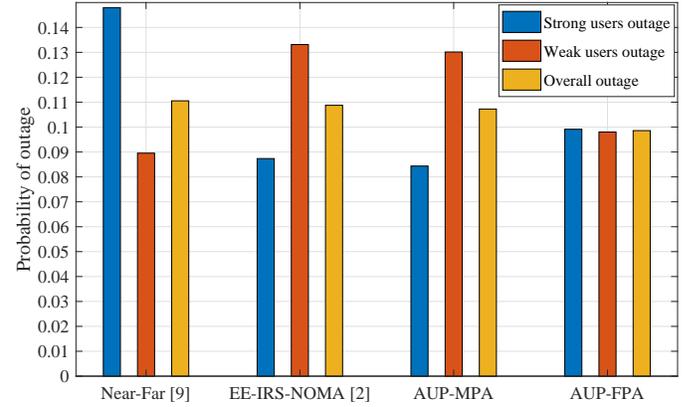}
    \caption{Probability of outage with various algorithms.}
    \label{fig:eps}
\end{figure}

We have also evaluated the probability of outage and presented the results for the same in Fig.~\ref{fig:eps}. It can be observed that \gls{AUP} with \gls{FPA} has similar levels of outage for both strong and weak users. In \gls{AUP} with \gls{MPA}, the outage for weak users is higher as compared to the strong user. This is because \gls{MPA} allocates more power to the strong user to achieve higher \gls{ASR}, and hence, results in this unfairness. Further, in the evaluation of Near-Far~\cite{NearFar2}, we have assumed $\gls{a1}=\gls{a1}_\text{\tiny LB}$, and hence, it results in a higher outage for the strong user. Both Near-Far and \gls{EE}-\gls{IRS}-\gls{NOMA} algorithms have higher overall outage than the \gls{AUP} algorithms. As shown in Fig.~\ref{fig:cdf},~\ref{fig:eps}, the proposed \gls{AUP} with \gls{MPA} and \gls{FPA} outperform the state-of-the-art algorithms and provide trade-offs between maximum \gls{ASR} and fairness. 

\section{Conclusion}
\label{sec:Conclusion}
In this letter, we have derived bounds on the imperfection in the phase compensation and the power allocation factors for  \gls{IRS}-assisted \gls{NOMA} systems. Using these bounds, we have proposed an adaptive user pairing algorithm to improve the achievable data rates of each user. We have then proposed power allocation algorithms to achieve maximum sum rate or ensure fairness, respectively. Through extensive simulations, we have shown that the proposed power allocation algorithms offer trade-offs between the achievable data rates and the fairness. In future, we plan to implement and validate the proposed algorithms on hardware testbeds.
\vspace{-0.2100cm}
\bibliographystyle{ieeetran}
\bibliography{Bibfile.bib}
\end{document}